\documentclass[12pt]{article}
\usepackage{epsfig}
\usepackage{amsfonts}
\usepackage{amsmath}
\usepackage{amssymb}
\usepackage{cite}
\begin{document}
\sffamily
\title{A method for extracting the resonance parameters from experimental cross
section}
\author{
S. A. Rakityansky$^{1)}$\footnote{\small e-mail: {\sf rakitsa@up.ac.za}}
\ and
\ N. Elander$^{2)}$\\
$^{1)}${\small\it Dept. of Physics, University of Pretoria, Pretoria 0002,
South Africa}\\
$^{2)}${\parbox{12cm}{\small\it Div.of Chemical Physics,
        Dept. of Physics, Stockholm University,\\
        Stockholm, SE-106 91, Sweden}}
}
\maketitle
\begin{abstract}
\noindent
The matrix elements of the multi-channel
Jost matrices are written in such a
way that their dependencies on all possible odd powers of channel momenta are factorized explicitly. As
a result the branching of the Riemann energy surface at all the channel
thresholds is represented in them via exact analytic expressions. The remaining
single-valued functions of the energy are expanded in the Taylor series near an
arbitrary point on the real axis. Using the thus obtained Jost matrices, the
$S$-matrix is constructed and then the
scattering cross section is calculated,
which therefore depends on the Taylor expansion coefficients. These coefficients
are considered as the adjustable parameters that are optimized to fit a given
set of experimental data. After finding the coefficients, the resonances are
located as zeros of the Jost matrix determinant at complex
energies. Within
this approach the $S$-matrix has proper analytic structure. This enables us not only to locate multi-channel resonances but also
to reproduce their partial widths as well as the
scattering cross section in the channels for which the data are not available.
\end{abstract}
\vspace{.5cm}
\noindent
PACS number(s): {03.65.Nk, 03.65.Ge, 24.30.Gd}\\[5mm]
\section{Introduction}
\label{sec.Introduction}
Quantum resonances play an important role in theoretical description and
intuitive understanding of physical processes taking place in the microscopic
world of molecules, atoms, nuclei, and various nano-systems. This is why
locating the corresponding spectral points $\mathcal{E}=E_r-\frac{i}{2}\Gamma$,
i.e. determining the resonance parameters, $E_r$ (resonance energy) and $\Gamma$
(resonance width), is an indispensable part of any theoretical modelling as
well as of an analysis of scattering data.\\

From the theoretical point of view, the resonances, being complex-valued
eigen-energies of the Hamiltonian, are not easy to locate. For each angular momentum, they appear as regular
strings on the so called non-physical sheet of the complex energy surface
(see, for example, Ref.~\cite{Kukulin}). The influence of these resonances on the corresponding
partial-wave cross section depends not only on their distribution over the energy
surface but also on the corresponding $S$-matrix
residues~\cite{r:ElRakit04,r:KsenIJQC09,r:jpb09,ElandRakit11}.\\

The analysis of scattering data, i.e. phenomenological extraction of resonance parameters, is perhaps even more difficult than their theoretical
determination. Indeed, a measured cross section is the sum of all partial-wave
cross sections, where the effects of all possible resonances add together. Therefore the sharp structures corresponding to individual resonances, may overlap and smear out their discrete character. Such an overlap very often happens even with resonances belonging to the same partial wave.\\

In a phenomenological search for resonances, the first step usually consists in performing the so called phase-shift analysis~\cite{Nichitiu1,Nichitiu2,Kukulin}, i.e. in obtaining numerical values of the partial-wave phase shifts or the corresponding reaction amplitudes, for a set of collision energies. Then the amplitudes known on the real axis, are analytically continued onto the complex-energy Riemann surface, where the resonances are identified as their singularities. Various methods differ in the way such an analytic continuation is done~\cite{Kukulin}.\\

The basic idea underlying the majority of existing approaches of this type, consists in fitting the experimental points with curves obtained from a phenomenological scattering amplitude, where possible resonance poles are artificially embedded by hand and their complex energies serve as the fitting parameters. The
most simple of such approaches is the Breit-Wigner parametrization~\cite{r:BW36}. Another version of the same approach
is based on the Fano parametrization~\cite{r:Fanoprof}, where the amplitude is
split in the resonant and background parts, which makes it more realistic and
allows one to treat more complicated "zigzags" of the cross section. In nuclear and particle physics, the resonances are usually introduced into the fitting procedure via a model propagator with explicit singularities at complex energies (unitary isobar model)~\cite{Lee,Tiator}.\\

Many authors emphasize (see, for example, Chapter 6 of the book \cite{Kukulin} and references therein as well as a more recent publication \cite{Svarc}) that for a reliable analytic continuation of the scattering amplitude (or $S$-matrix) it is very important to construct its phenomenological expression preserving proper analytic structure, i.e. its branching points, various symmetries etc. This can easily be achieved for the single-channel problems and with some effort for two-channel ones (Dalitz-Tuan representation)~\cite{Badalyan}. However, when the number of channels is three or more, constructing the $S$-matrix with correct branching at all the thresholds becomes extremely difficult within the traditional approaches~\cite{Badalyan,Surovtsev}.\\

In the present paper, we resolved this difficulty. For obtaining the $S$-matrix, we use the structure of the Jost matrix derived in our early publications~\cite{my2009,my2011}, where all the factors responsible for the branching are given analytically for an arbitrary number of channels. This enables us to perform the analytic continuation and the search for resonances in a model-independent way, i.e. without embedding the resonance poles by hand.\\

\section{Jost matrices}
\label{sec.Jostmatrices}
Consider an $N$-channel collision process that can symbolically be written as a
chemical or nuclear reaction,
$$
     A+B\ \rightarrow\ \left\{
     \begin{array}{cc}
     A+B & ,\\
     C+D & ,\\
     \dots\dots & ,
     \end{array}\right.
$$
with two-body systems in both the initial and final states. We assume that the
interaction forces for these systems are of a short-range type, i.e. they vanish
at large distances ($r\to\infty$) faster than the Coulomb force. The wave
function describing such a process at the collision energy $E$ is a column
matrix,
$$
     \Psi(E,\vec{r})=
     \begin{pmatrix}
     \psi_1(E,\vec{r})\\
     \psi_2(E,\vec{r})\\
     \dots
     \end{pmatrix}\ ,
$$
where each of the $N$ lines corresponds to a separate channel. The channels are
characterized not only by the type of the particles ($A+B$, $C+D$, etc.) but also
by the complete sets of the corresponding quantum numbers (such as the threshold
energy, angular momentum, spin, etc.). In other words, any two states
of the system that differ by at least one quantum number are considered as
different channels even if the type of the particles and the threshold energies
are the same. This means that a channel state has a definite value of the
angular momentum and thus its angular dependence can be factorized as
$$
    \psi_n(E,\vec{r})=\frac{u_n(E,r)}{r}Y_{\ell_nm_m}(\theta,\varphi)\ .
$$
The set of radial wave functions $u_n$ obey the coupled system of radial
Schr\"odinger equations,
\begin{equation}
\label{mult_chan_shreq}
    \left[\partial^2_r+k_n^2-\frac{\ell_n(\ell_n+1)}{r^2}\right]
    u_{n}(E,{r})
    =
    \frac{2\mu_n}{\hbar^2}
    \sum_{n'=1}^NV_{nn'}(r)u_{n'}(E,{r})\ ,
\end{equation}
where the coupling is due to the off-diagonal elements of the
interaction-potential matrix $V_{nn'}$. In Eq. (\ref{mult_chan_shreq}), the
channel momentum,
\begin{equation}
\label{chmom}
    k_n=\sqrt{\frac{2\mu_n}{\hbar^2}(E-E_n)}\ ,
\end{equation}
is determined by the difference between the total energy $E$ and the channel
threshold $E_n$, as well as by the reduced mass $\mu_n$ in the channel $n$.\\

Each of the $N$ equations of the set (\ref{mult_chan_shreq}) is of the second
order. In the theory of differential equations (see, for example, Ref.
\cite{brand}) it is shown that such a set has $2N$ linearly independent
solutions, i.e. $2N$ linearly independent columns that form a basis in the
solution space, and only half of these columns are regular at $r=0$. The
regular columns can be combined in a square matrix $\Phi(E,r)$, which is called
the fundamental matrix of regular solutions. Since a physical wave function
must be regular, it is a linear combination of the columns of the matrix
$\Phi(E,r)$.\\

When the particles move away from each other ($r\to\infty$), the potential
matrix tends to zero and thus the right hand side of Eq. (\ref{mult_chan_shreq})
vanishes. The remaining set of $N$ second order (Riccati-Bessel) equations,
\begin{equation}
\label{mult_chan_shreq_ass}
    \left[\partial^2_r+k_n^2-\frac{\ell_n(\ell_n+1)}{r^2}\right]
    u_{n}(E,{r})
    \approx 0\ ,
    \qquad r\to\infty\ ,
\end{equation}
has $2N$ linearly independent column-solutions. These $2N$ columns can be
combined in two diagonal square matrices,
\begin{equation}
\label{inwaves}
    W^{\rm (in)}=
    \begin{pmatrix}
    h^{(-)}_{\ell_1}(k_1r) & 0 & \cdots & 0\\
    0 & h^{(-)}_{\ell_2}(k_2r) & \cdots & 0\\
    \vdots & \vdots & \vdots & \vdots \\
    0 & 0 & \vdots & h^{(-)}_{\ell_N}(k_Nr)\\
    \end{pmatrix}
\end{equation}
\begin{equation}
\label{outwaves}
    W^{\rm (out)}=
    \begin{pmatrix}
    h^{(+)}_{\ell_1}(k_1r) & 0 & \cdots & 0\\
    0 & h^{(+)}_{\ell_2}(k_2r) & \cdots & 0\\
    \vdots & \vdots & \vdots & \vdots \\
    0 & 0 & \vdots & h^{(+)}_{\ell_N}(k_Nr)\\
    \end{pmatrix}
\end{equation}
that involve the Riccati-Hankel functions $h_{\ell}^{(\pm)}(kr)$ and
represent the {\it in}-coming and {\it out}-going spherical waves in all $N$
channels.\\

The $2N$ columns of the matrices (\ref{inwaves}) and (\ref{outwaves})
constitute a basis in the solution space at large distances and thus each
column of the fundamental matrix $\Phi(E,r)$ is their linear combination when
$r\to\infty$. This can be written as
\begin{equation}
\label{phi_ass}
    \Phi(E,r)
    \ \mathop{\longrightarrow}\limits_{r\to\infty}
    \ W^{\rm (in)}(E,r)F^{\rm (in)}(E)+
      W^{\rm (out)}(E,r)F^{\rm (out)}(E)\ ,
\end{equation}
where the combination coefficients are combined in the square matrices
$F^{\rm (in)}(E)$ and $F^{\rm (out)}(E)$. They are functions of the energy and
are called the Jost matrices.\\

The $S$-matrix that completely determines all the scattering observables, is
expressed via the Jost matrices (the details can be found, for example, in the
textbook \cite{Taylorbook} or in the papers
\cite{r:ElRakit04,my04,my06,my2011}),
\begin{equation}
\label{S_matrix}
    S(E)=F^{\rm (out)}(E)\left[F^{\rm (in)}(E)\right]^{-1}\ .
\end{equation}
The resonances are those spectral points
\begin{equation}
\label{Res_energy}
   {\cal E}=E_r-\frac{i}{2}\Gamma\ ,\qquad
   E_r>0\ ,\quad \Gamma>0\ ,
\end{equation}
on the Riemann surface of the energy, where
\begin{equation}
\label{spectral}
    \det F^{\rm (in)}({\cal E})=0\ .
\end{equation}
The energy surface has a square-root branching point at every channel threshold
$E_n$. This is because the Jost matrices depend on the energy $E$ via the
channel momenta (\ref{chmom}) and for each of them there are two
possible choices of the sign in front of the square root. The resonance
spectral points are located on the so called non-physical sheet of this Riemann
surface, i.e. such a layer of the surface where all the channel momenta have
negative imaginary parts. In the numerical calculations, the choice of the
sheet is done by an appropriate choice of the signs in front of the square
roots (\ref{chmom}).

\section{Analytic structure of the Jost matrices}
\label{sec.Analytic_structure}
In the present paper, our main goal is to find a way of a reliable
parametrization of experimental data, such that it would allow us to locate the
resonance spectral points at complex energies. As we described in the previous
section, the multi-channel $S$-matrix has very complicated
energy dependence
via the channel momenta and therefore is defined on a Riemann surface with an
intricate connection of many layers. This means that a straightforward
parametrization of such a matrix using an arbitrarily chosen functional form
may give erroneous results. When choosing the parametrization form, it is
important to take into account as much information on the symmetry properties
and analytic structure of the $S$-matrix as possible.\\

First of all, we notice that the
 $S$-matrix is a kind of ``ratio", given by Eq.
(\ref{S_matrix}), of the Jost matrices $F^{\rm (out)}$ and
$F^{\rm (in)}$,
which are not completely independent of each other but rather are somehow related.
In other words, the parameters in the ``numerator'' and ``denominator" of
(\ref{S_matrix}) should be the same. Indeed, as we found
in Ref. \cite{my2011}, for the systems interacting via short-range
potentials, the Jost matrices have the following structure:
\begin{eqnarray}
\label{fin_semi}
    F^{\rm (in)}_{mn}(E) &=&
                 \frac{k_n^{\ell_n+1}}{2k_m^{\ell_m+1}}{A}_{mn}(E)-
                 \frac{ik_m^{\ell_m}k_n^{\ell_n+1}}{2}{B}_{mn}(E)\ ,\\[3mm]
\label{fout_semi}
    F^{\rm (out)}_{mn}(E) &=&
                 \frac{k_n^{\ell_n+1}}{2k_m^{\ell_m+1}}{A}_{mn}(E)+
                 \frac{ik_m^{\ell_m}k_n^{\ell_n+1}}{2}{B}_{mn}(E)\ ,
\end{eqnarray}
where the energy dependent matrices $A(E)$ and $B(E)$ are the same for both $F^{\rm (in)}$ and $F^{\rm (out)}$. Moreover,
in the same Ref. \cite{my2011} it was established that the matrices $A(E)$ and
$B(E)$ are single-valued analytic functions of the energy defined on a single
one-layer energy plane. In other words, the matrices $A(E)$ and $B(E)$ are the
same for all the sheets of the Riemann surface and all the complications
stemming from the branching points are isolated in Eqs. (\ref{fin_semi})
and (\ref{fout_semi}) via the explicit factors depending on the channel momenta.
Therefore, by using the expressions (\ref{fin_semi}) and (\ref{fout_semi}) in
Eq. (\ref{S_matrix}), we guarantee that the ``numerator" is properly related to
the ``denominator" (which means guaranteed unitarity on the real axis) and that
all the branching points are properly embedded. If we find an adequate
parametrization of the matrices $A(E)$ and $B(E)$, the resulting
$S$-matrix
will automatically have correct values on all the sheets of the Riemann
surface.\\

Since the matrices $A(E)$ and $B(E)$ are analytic functions of the variable
$E$, they can be expanded in the Taylor series,
\begin{equation}
\label{PSEAB}
     A(E) = \sum_{n=0}^\infty(E-E_0)^na_n(E_0)\ ,\qquad
     B(E) = \sum_{n=0}^\infty(E-E_0)^nb_n(E_0)\ ,
\end{equation}
near an arbitrary point $E_0$ within the domain of their analyticity. Here the
expansion coefficients $a_n$  and $b_n$ are matrices of the same dimension as
$A$ and $B$. They depend on the choice of the point $E_0$.\\

In Ref. \cite{my2011} it was shown that for a given potential the expansion
coefficients $a_n$ and $b_n$ can be obtained as the asymptotic values of the
solutions of certain set of differential equations. In the present paper we are
not going to calculate these coefficients since we assume that the potential is
not known. Instead, we will use $a_n$ and $b_n$ as fitting parameters in order
to reproduce experimental scattering cross section. As soon as the optimal
expansion coefficients corresponding to the experimental data are found, we can
use them to obtain the matrices $A(E)$ and $B(E)$ and through them the
Jost
matrix $F^{\rm (in)}(E)$ given by Eq. (\ref{fin_semi}).
 Then we can locate the
resonances as the complex roots of Eq. (\ref{spectral}).\\

Although we are not going to calculate the expansion coefficients, the fact
that they are solutions of certain differential equations derived in Ref.
\cite{my2011} is important. Indeed, that differential equations have real
boundary conditions and all their coefficients are real if $E_0$ is chosen on the
real axis. This means that for real $E_0$ the matrices $a_n$ and $b_n$ do not
have imaginary parts, i.e. the total number of fitting parameters is $2(M+1)N^2$
where $N$ is the matrix dimension (number of channels) and $M$ is the highest
power in the expansions
\begin{equation}
\label{approxAB}
     A(E) \approx \sum_{n=0}^M(E-E_0)^na_n(E_0)\ ,\qquad
     B(E) \approx \sum_{n=0}^M(E-E_0)^nb_n(E_0)\ .
\end{equation}
If the central point $E_0$ of the expansion is taken not on the real axis, then
the number of the parameters we have to fit is doubled.\\

The approximate expansions (\ref{approxAB}) can only be accurate within a circle
around $E_0$. In Ref. \cite{my2011} it is demonstrated how the radius of such a
circle increases with the number of terms taken into account. When using this
approach to extract the resonance parameters from experimental cross section,
we therefore should not fit the data within a too wide energy-interval. The
central point $E_0$ must be placed somewhere in the middle of the interval
where it is expected to find a resonance. An adequate width of the interval
around $E_0$ unfortunately cannot be estimated. It is necessary to repeat the
analysis with several different widths and make the decision on the basis of
the stability of the results thus obtained.\\

\section{Fitting procedure}
\label{sec.Fitting_procedure}
Suppose we have sets of experimental data for several channels,
$m\to n$,  $m'\to n'$, etc.:
$$
   \begin{array}{rcl}
   \sigma_{mn}\left(E^{(mn)}_i\right) &\pm& \delta^{(mn)}_i\ ,\\[3mm]
   \sigma_{m'n'}\left(E^{(m'n')}_j\right) &\pm& \delta^{(m'n')}_j\ ,\\[3mm]
   \cdots\cdots\cdots\cdots && \text{etc.}
   \end{array}
   \qquad
   \begin{array}{rcl}
   i &=& 1,2,3,\dots,N^{(mn)}\ ,\\[3mm]
   j &=& 1,2,3,\dots,N^{(m'n')}\ ,\\[3mm]
   && \cdots\cdots
   \end{array}
$$
Here $\sigma_{mn}$ and $\delta^{(mn)}$ are the cross section in the channel
$m\to n$ and the corresponding standard deviation while the energy interval
covered by the points $E^{(mn)}_i$ is around the energy $E_0$ where we expect to
find a resonance. In order to parametrize the Jost matrix,
 we construct the
$\chi^2$ function
\begin{eqnarray}
\label{chi2}
   \chi^2 &=&
   \displaystyle
   \sum_{i=1}^{N^{(mn)}}\left[\frac{\sigma_{mn}(E^{(mn)}_i)-
   \sigma^{\mathrm{fit}}_{mn}(E^{(mn)}_i)}
   {\delta^{(mn)}_i}\right]^2\\[3mm]
\nonumber
   &+&
   \displaystyle
   \sum_{j=1}^{N^{(m'n')}}\left[\frac{\sigma_{m'n'}(E^{(m'n')}_j)-
   \sigma^{\mathrm{fit}}_{m'n'}(E^{(m'n')}_j)}
   {\delta^{(m'n')}_j}\right]^2
   \ +\ \cdots\ \text{etc.}\ ,
\end{eqnarray}
where the fitting cross section for the channel $m\to n$
\begin{equation}
\label{cross_section_sigma_nm}
   \sigma^{\mathrm{fit}}_{mn}(E)=
   \frac{\pi}{k_m^2}(2\ell_m+1)\left|S_{nm}(E)-\delta_{nm}\right|^2
\end{equation}
depends on the expansion coefficients of (\ref{approxAB}) via Eqs.
(\ref{fin_semi}), (\ref{fout_semi}), and (\ref{S_matrix}). These coefficients
therefore serve as the fitting parameters.\\

The time-reversal invariance leads to the so called detailed balance theorem
which means that the $S$-matrix is symmetric with
respect to the transposition,
i.e. $S_{mn}=S_{nm}$. If we simply minimize the $\chi^2$-function given by Eq.
(\ref{chi2}), this symmetry is not guaranteed. If $N$ is the number of channels (the dimension of the matrices), then the symmetry gives us $(N^2-N)/2$ equations (the number of the elements above the diagonal of the matrix) relating the variational parameters. By solving this set of equations, we can reduce the number of such parameters. Although the equations are not simple, this can always be done numerically for any reasonable value of $N$.\\

There is a more simple way of making the $S$-matrix symmetric although it requires to vary a bit more parameters.
A set of the optimal parameters
$a_0,a_1,\dots,a_M,b_0,b_1,\dots,b_M$
that give a symmetric $S$-matrix, can be obtained by minimizing the
generalized $\chi^2$-function
\begin{equation}
\label{goalFunct}
   \mathcal{X}^2(a_0,a_1,\dots,a_M,b_0,b_1,\dots,b_M)=
   \chi^2+\sum_{m<n,j}\left|S_{mn}(E_j)-S_{nm}(E_j)\right|^2\ ,
\end{equation}
where at all experimental points the differences between the off-diagonal elements are included.\\

After finding the optimal parameters, we obtain analytic expressions for the Jost matrices and the $S$-matrix, valid within a circle around $E_0$ on all the sheets of the Riemann surface. Using these expressions, we should be able not only to locate the nearest
resonances but also to calculate the cross sections in all the other channels for
which we do not have experimental data.

\section{Examples}
\label{sec.Examples}
The proposed procedure for parametrizing experimental cross section and
thus locating the resonances needs to be demonstrated by a couple of
simple and clear examples. In such examples, we should know the resonance
parameters beforehand. This will give us the feeling of how accurate the
procedure is.\\

As the examples, we chose the well-known and well studied one- and two-channel
models specified by certain potentials (see the next two sections). We use these
potentials to artificially generate ``experimental'' data points around the
energies where they support resonances, and then try to recover these resonances
using the suggested
parametrization method. Since the exact values of the resonance parameters are
known, this shows us how reliable the proposed method is.

\subsection{Single-channel model}
\label{sec.single}
\begin{figure}[ht!]
\centerline{\epsfig{file=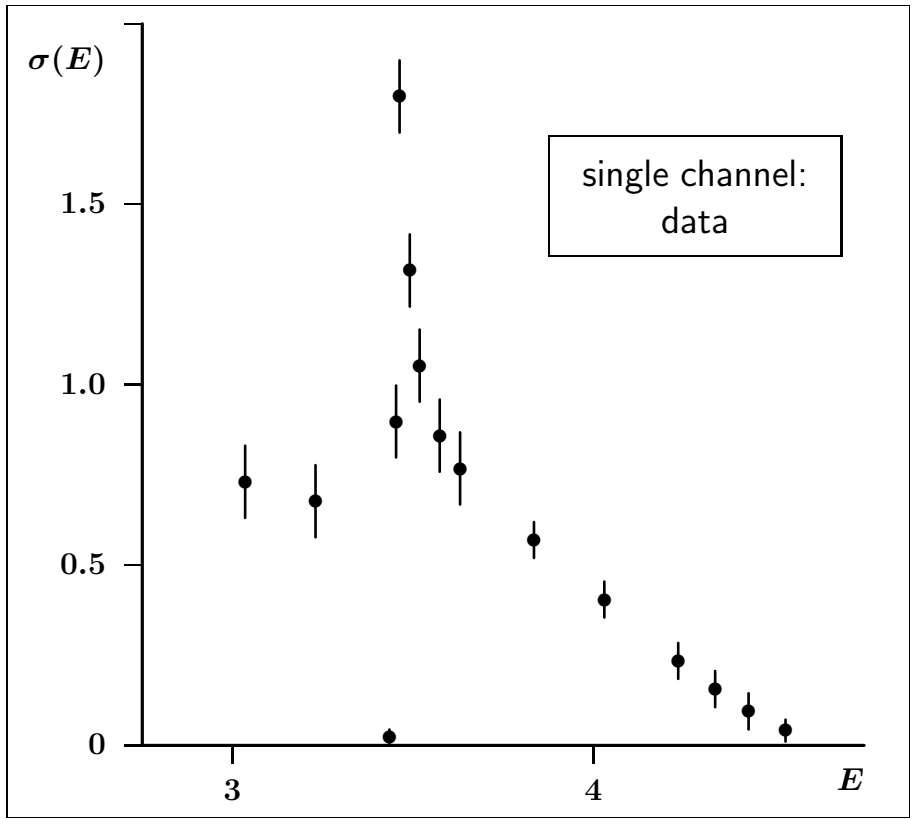}}
\caption{\sf
Artificial data points for the single-channel model ~\protect(\ref{1channelV}).
}
\label{fig.1channel_data}
\end{figure}

\begin{figure}[ht!]
\centerline{\epsfig{file=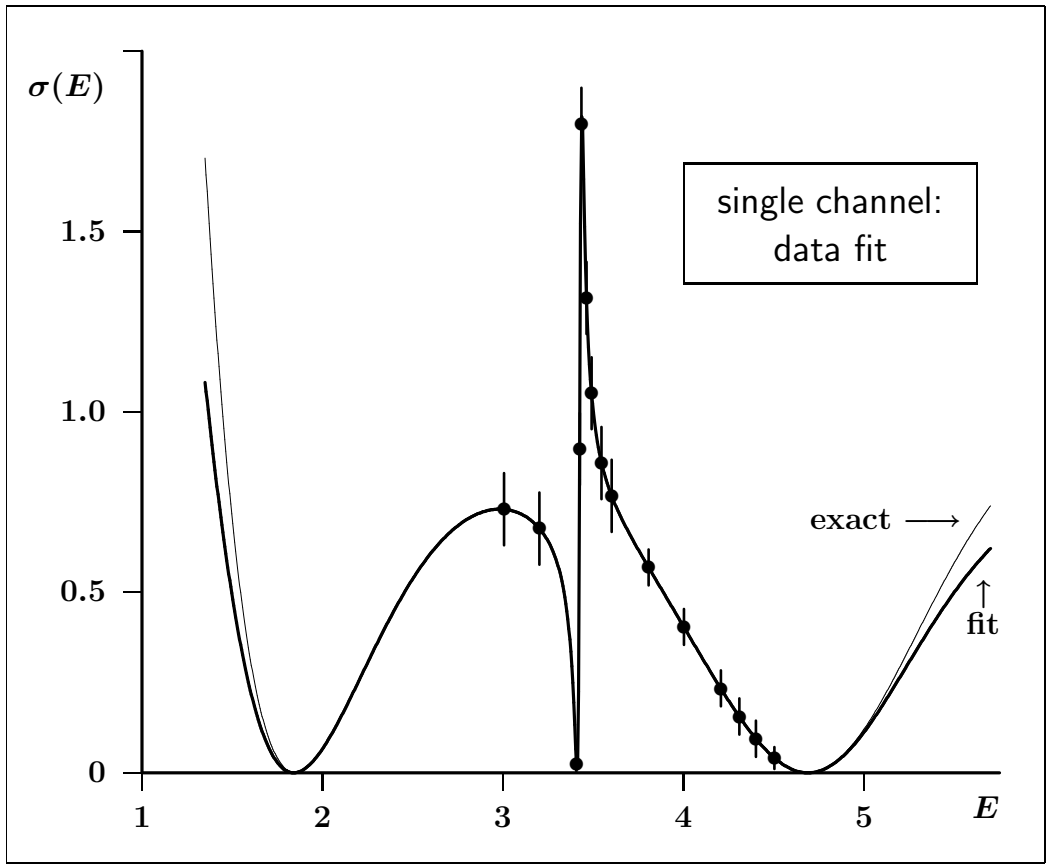}}
\caption{\sf
Exact cross section (thin curve) for the single-channel model
~\protect(\ref{1channelV}) and the result of fitting of the data (thick curve)
with $M=5$ and $E_0=3.4$ in Eq.~\protect(\ref{approxAB}).
}
\label{fig.1channel_fit}
\end{figure}

The simple potential barrier given by
\begin{equation}
\label{1channelV}
    V(r) = 7.5r^2e^{-r}\ ,
\end{equation}
is very often used as a testing ground for new theoretical
methods\cite{bainpot}. In this
model, the units are such that $\hbar^2/\mu=1$ and thus the energy as well as
the distances are dimensionless. It has a rich spectrum of resonances. The first
two of them with $\ell=0$ are~\cite{my02}:
\begin{eqnarray}
\label{single_exactresonance_1}
   \mathcal{E}_1^{\mathrm{exact}} &=& 3.426390-\frac{i}{2}0.025549\ ,\\
\label{single_exactresonance_2}
   \mathcal{E}_2^{\mathrm{exact}} &=& 4.834807-\frac{i}{2}2.235753\ .
\end{eqnarray}
Let us assume that we are given experimental $S$-wave scattering cross section
for a single-channel system in the energy interval $3<E<4.5$ as is shown in Fig.
\ref{fig.1channel_data}. Actually, these 15 points are generated using the
potential (\ref{1channelV}), along the corresponding exact cross section shown
in Fig. \ref{fig.1channel_fit} (thin curve), which has a sharp zigzag near the
first very narrow resonance (\ref{single_exactresonance_1}).\\

In order to fit the data, we use the approximate expressions (\ref{approxAB})
with $E_0=3.4$ and $M=5$. Being substituted into Eqs. (\ref{fin_semi}) and
(\ref{fout_semi}), they give us the Jost functions, from which we obtain the
$S$-matrix (\ref{S_matrix}) and finally the cross section
(\ref{cross_section_sigma_nm}). As we mentioned before, for a real $E_0$ the
coefficients (fitting parameters) $a_n$ and $b_n$ are also real. This means that
we have to adjust 12 parameters that minimize the $\chi^2$-function of the
type (\ref{chi2}) where only single elastic channel is taken into account.\\

As the minimization tool, we used well known program ``MINUIT'' from the CERN
library~\cite{minuit1,minuit2}. In order to avoid the situation when we stuck in a local minimum, we
repeated the minimization procedure several hundreds of times with randomly
chosen initial values of the parameters and took the best findings. The best
minimum we found was with $\chi^2=1.5\times10^{-6}$. The result of this fitting
is shown in Fig. \ref{fig.1channel_fit} (thick curve). There is no visible
difference between the exact (thin curve) and fitted curve not only within the
interval $3<E<4.5$ covered by the ``experimental'' points, but also at the
nearby points to the left and to the right of that interval. This is because we
use proper analytic structure of the fitting $S$-matrix.\\

With thus found set of parameters, we located two zeros (\ref{spectral}) of the
function (\ref{fin_semi}) nearest to the real axis. They gave us the approximate
(recovered from experimental data) resonance energies:
\begin{eqnarray}
\label{single_approxresonance_1}
      \mathcal{E}^{\mathrm{fit}}_1 &=& 3.426388-\frac{i}{2}0.025531\ ,\\
\label{single_approxresonance_2}
      \mathcal{E}^{\mathrm{fit}}_2 &=& 4.821657-\frac{i}{2}2.036732\ .
\end{eqnarray}
Comparing them with the corresponding exact values
(\ref{single_exactresonance_1}) and (\ref{single_exactresonance_2}),
we see that the result of fitting is very accurate for the
$S$-matrix not only
on the real axis but also in the nearby domain of the complex $E$-surface.
Actually, when choosing the ``experimental'' points we did not intend to
reproduce the second resonance which is rather far away from the real axis.\\

In addition to finding the optimal values of the parameters, the minimization
program ``MINUIT'' provides statistical errors (standard deviations) for them.
This is done by calculating the matrix of partial derivatives of the minimized
function with respect to all the parameters~\cite{minuit1,minuit2}. Using these statistical errors, we
can estimate the corresponding errors of the resonance parameters we found.\\

To this end we considered the optimal values of the parameters as their mean
values and randomly varied all the parameters around these values using a
random-number generator with the Gaussian distribution of the width equal to the statistical errors. For each random choice of the parameters, we located zeros
of the Jost function and then calculated their mean values and the standard
deviations. After 1000 variations, we obtained:
\begin{eqnarray}
\label{single_statresonance_1}
      \mathcal{E}^{\mathrm{fit}}_1 &=&
      \left(3.426309 \pm 0.005361\right)
      -\frac{i}{2}
      \left(0.025921\pm 0.010444\right)\ ,\\
\label{single_statresonance_2}
      \mathcal{E}^{\mathrm{fit}}_2 &=&
      \left(3.810114 \pm 0.448626\right)
      -\frac{i}{2}
      \left(1.457056\pm 1.090140\right)\ .
\end{eqnarray}
As one would expect, the reliability of the recovery of the second resonance is
not that good as for the first one (around which the ``experimental'' points
were taken). We however did not even expected to recover the second resonance at
all.

\subsection{Two-channel model}
\label{sec.double}
The two-channel potential,
\begin{equation}
\label{2channelV}
     V(r) = \begin{pmatrix}
    -1.0 & -7.5\\
    -7.5 & 7.5\\
    \end{pmatrix}
    r^2e^{-r}\ ,
\end{equation}
of famous Noro and Taylor model \cite{norotaylor} extends the single-channel
potential of Sec. \ref{sec.single}. It is written in the same dimensionless
units with equal reduced masses $\mu_1=\mu_2$ and angular momenta
$\ell_1=\ell_2=0$ in both channels. The threshold energies for the channels are
$E_1=0$ and $E_2=0.1$.\\

The first three resonances of the Noro-Taylor model are given in Table
\ref{table.2spectrum} (the calculations can be found, for example, in Ref.~\cite{r:ElRakit04}). The other resonances supported by the potential
(\ref{2channelV}), are too wide and therefore too far from the real axis.\\

\begin{figure}[ht!]
\centerline{\epsfig{file=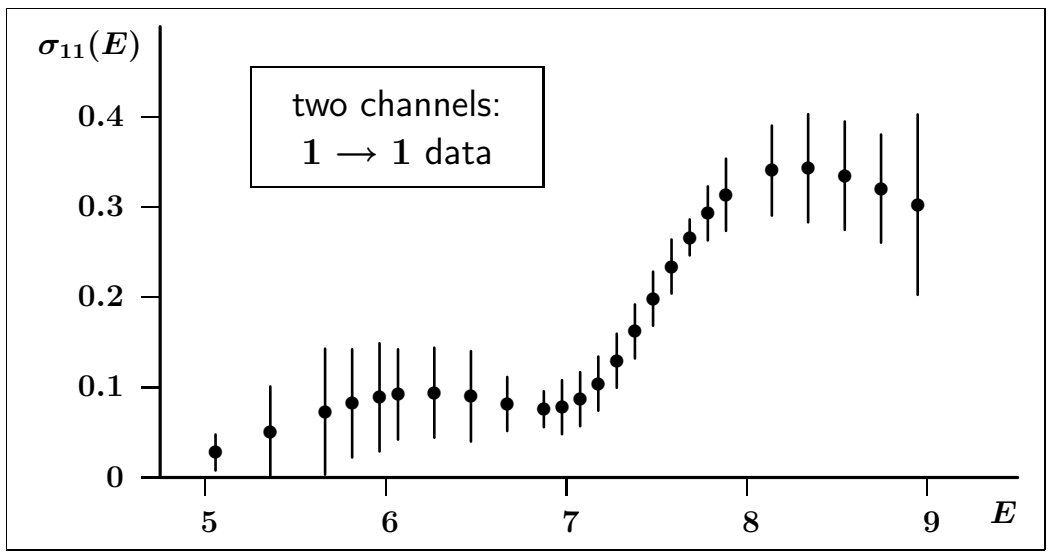}}
\caption{\sf
Artificial data points for the first elastic channel of the model
~\protect(\ref{2channelV}).
}
\label{fig.2channel11_data}
\end{figure}

\begin{figure}[ht!]
\centerline{\epsfig{file=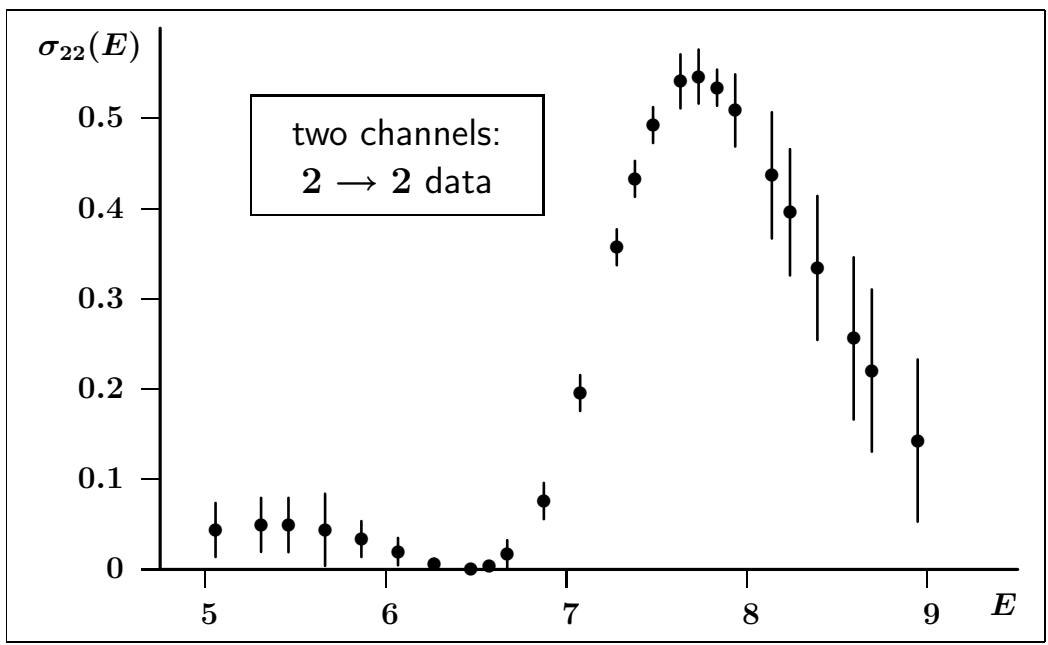}}
\caption{\sf
Artificial data points for the second elastic channel of the model
~\protect(\ref{2channelV}).
}
\label{fig.2channel22_data}
\end{figure}

Since in the previous section we already demonstrated how the method works for a
narrow resonance, considering the two-channel model, we focus our attention on
the second resonance of the Table \ref{table.2spectrum}, which is rather wide.
We took ``experimental'' data in the elastic channels $1\to1$ and $2\to2$ within
the energy interval $5<E<9$. If one considers this segment of the real axis as
the diameter of a circle in the complex plane, then such a circle will include
the resonance point which we are looking for. The artificial data points (25 in
the $1\to1$ and 25 in the $2\to2$ channels) are shown in Figs.
\ref{fig.2channel11_data} and \ref{fig.2channel22_data}.\\

In the same way as for the single-channel model, these 50 data points were
fitted using the approximate matrices (\ref{approxAB}) with $M=5$ and real
$E_0=7.25$. In such a case the matrices (\ref{approxAB}) are real and in total
we have to adjust 48 parameters (matrix elements of $a_n$ and $b_n$) in order to
minimize the generalized $\chi^2$-function (\ref{goalFunct}). After a thousand
attempts with randomly chosen initial values for $a_n$ and $b_n$,
the best value of the minimum we found was $\mathcal{X}^2=1.9\times10^{-4}$.\\

\begin{figure}[ht!]
\centerline{\epsfig{file=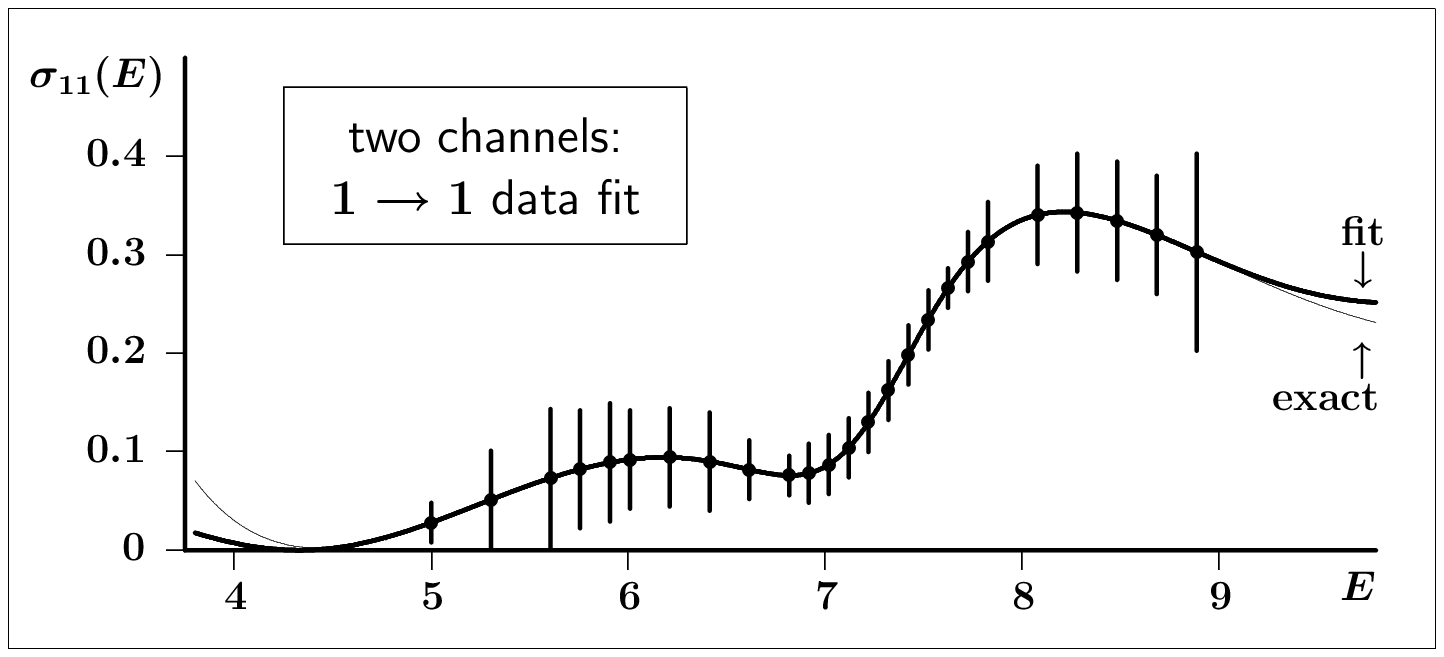}}
\caption{\sf
Exact elastic cross section $1\to1$ (thin curve) for the two-channel model
~\protect(\ref{2channelV}) and the result of fitting of the data (thick curve)
with $M=5$ and $E_0=7.25$ in Eq.~\protect(\ref{approxAB}).
}
\label{fig.2channel11_fit}
\end{figure}

\begin{figure}[ht!]
\centerline{\epsfig{file=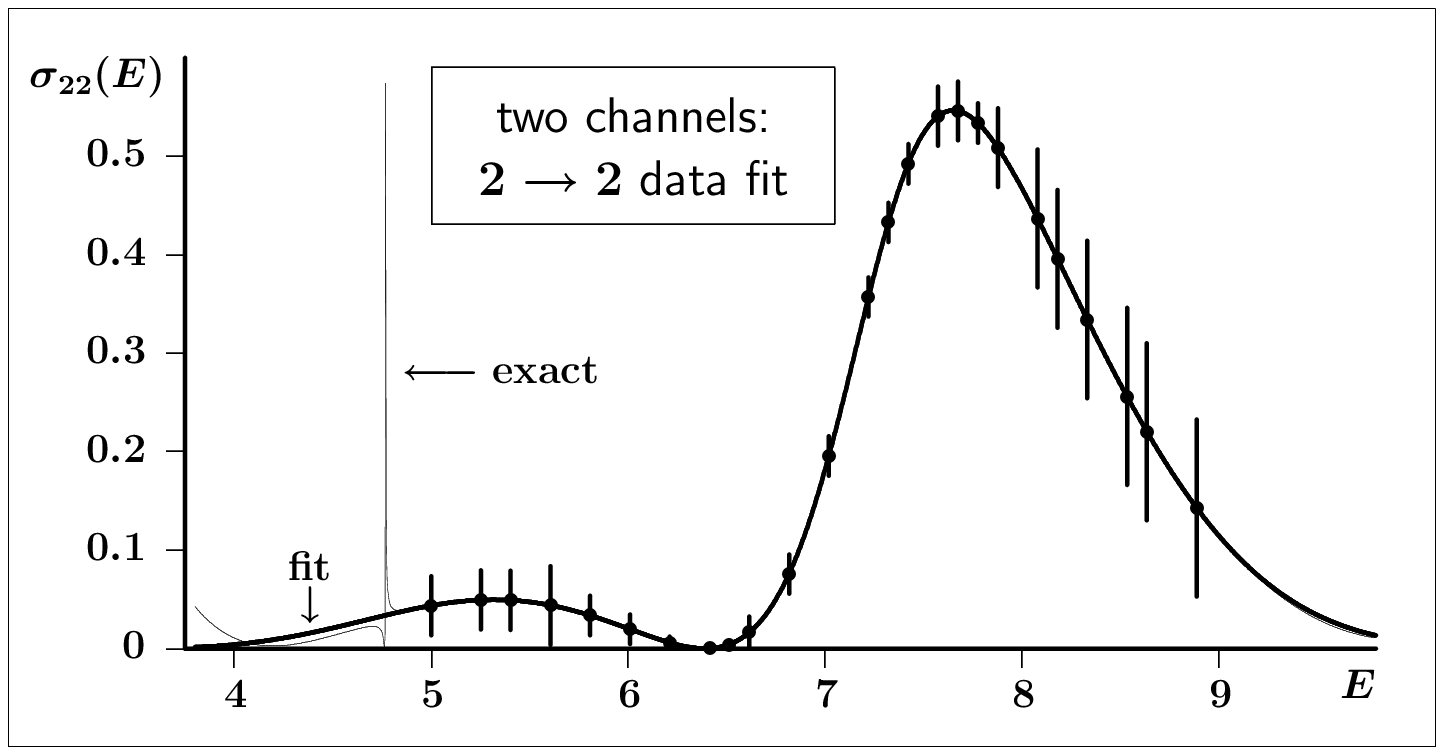}}
\caption{\sf
Exact elastic cross section $2\to2$ (thin curve) for the two-channel model
~\protect(\ref{2channelV}) and the result of fitting of the data (thick curve)
with $M=5$ and $E_0=7.25$ in Eq.~\protect(\ref{approxAB}).
}
\label{fig.2channel22_fit}
\end{figure}

The results of this fitting are shown in Figs. \ref{fig.2channel11_fit} and
\ref{fig.2channel22_fit} (thick curves). With rather small value of the
$\mathcal{X}^2$ achieved, there are no visible difference between the exact
(thin curves) and fitted curves within the interval covered by the
``experimental'' points. However the extremely sharp (first) resonance to the
left of this interval is missing since there are no data points reflecting it.\\

\begin{figure}[ht!]
\centerline{\epsfig{file=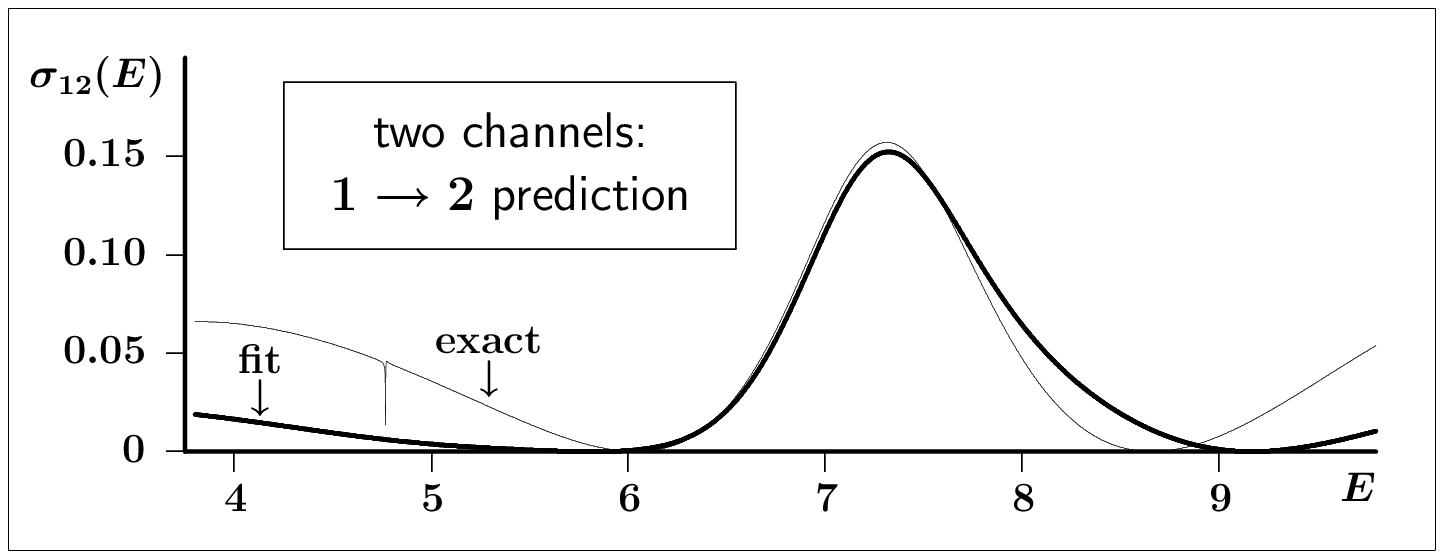}}
\caption{\sf
Exact inelastic cross section $1\to2$ (thin curve) for the two-channel model
~\protect(\ref{2channelV}) and the prediction (thick curve)
based on fitting of the data in the elastic channels $1\to1$ and $2\to2$ with
$M=5$ and $E_0=7.25$ in Eq.~\protect(\ref{approxAB}).
}
\label{fig.2channel12_fit}
\end{figure}

\begin{figure}[ht!]
\centerline{\epsfig{file=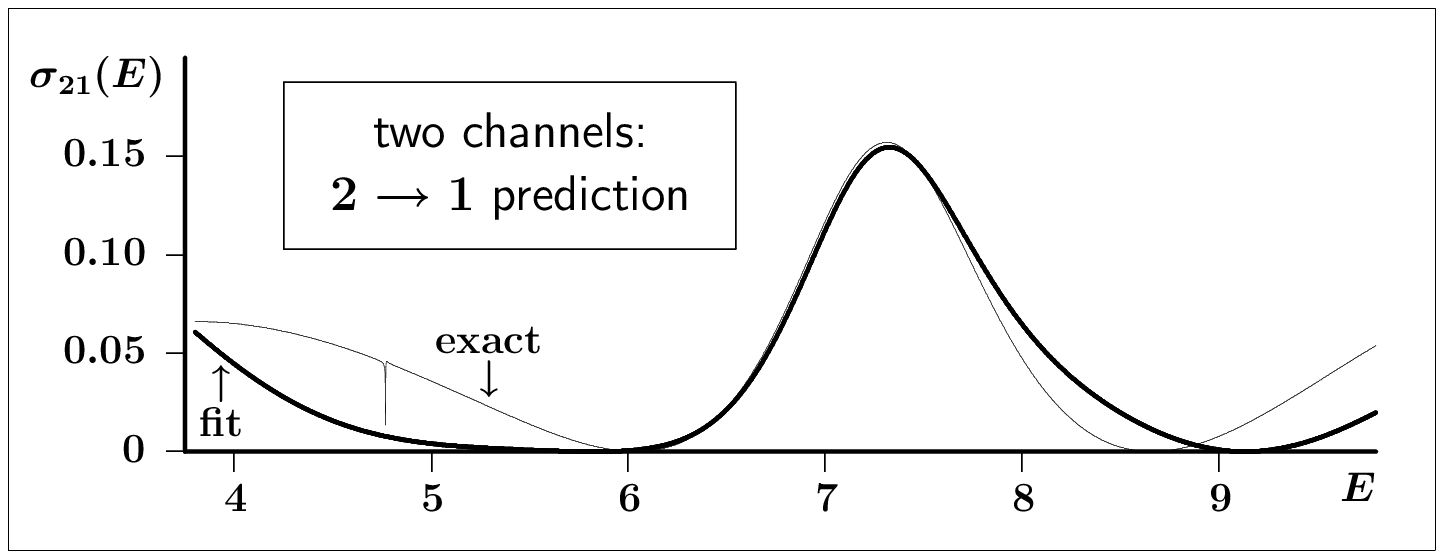}}
\caption{\sf
Exact inelastic cross section $2\to1$ (thin curve) for the two-channel model
~\protect(\ref{2channelV}) and the prediction (thick curve)
based on fitting of the data in the elastic channels $1\to1$ and $2\to2$ with
$M=5$ and $E_0=7.25$ in Eq.~\protect(\ref{approxAB}).
}
\label{fig.2channel21_fit}
\end{figure}

Using the same approximate (fitted) Jost matrices, we calculated the cross
sections for the transition processes $1\to2$ and $2\to1$, for which no data
points were taken. In Figs. \ref{fig.2channel12_fit} and
\ref{fig.2channel21_fit}, the comparison of the approximate (thick) and exact
(thin) curves shows that we are able to rather accurately predict the cross
section in one channel of the reaction on the basis of the data available in
the other channels.\\

Looking for the zero (nearest to $E_0$) of the determinant of the approximate Jost matrix, we found the following resonance:
\begin{eqnarray}
\label{double_approxresonance_2}
   \mathcal{E}_2^{\mathrm{fit}} &=& 7.250742-\frac{i}{2}1.513332\ ,
\end{eqnarray}
which is very close to its exact location (see the second line of Table
\ref{table.2spectrum}).\\

In order to find the partial widths, we use the method described in
Ref. \cite{r:ElRakit04}, where it was shown that the ratio of the partial
widhts can be found using the matrix elements of the Jost matrices, namely,
\begin{equation}
\label{GGratio}
   \frac{\Gamma_1}{\Gamma_2}=\left|
   \frac{F^{\rm (out)}_{11}F^{\rm (in)}_{22}-
         F^{\rm (out)}_{12}F^{\rm (in)}_{21}}
        {F^{\rm (out)}_{22}F^{\rm (in)}_{11}-
         F^{\rm (out)}_{21}F^{\rm (in)}_{12}}\right|_{E=\mathcal{E}}\ .
\end{equation}
Together with the fact that $\Gamma_1+\Gamma_2=\Gamma$, the knowledge of the
Jost matrices allows us to easily find $\Gamma_1$ and $\Gamma_2$.
For the result (\ref{double_approxresonance_2}), this gives
\begin{equation}
\label{G1G2}
      \Gamma_1^{\mathrm{fit}}=0.347732\ ,\qquad
      \Gamma_2^{\mathrm{fit}}=1.165600\ .
\end{equation}
These values reasonably well reproduce the corresponding exact partial widths
given in Table \ref{table.2spectrum}.\\

In the same way as for the single-channel case, we randomly varied (1000 times)
the optimal parameters $a_n$ and $b_n$, using the errors provided by the
``MINUIT'', and found the following mean values of the resonance energy and
widths together with the corresponding standard deviations:
\begin{equation}
\label{double_statresonance_2}
      \mathcal{E}^{\mathrm{fit}}_2 =
      \left(7.251593 \pm 0.292781\right)
      -\frac{i}{2}
      \left(1.085568\pm 0.537132\right)\ ,
\end{equation}
\begin{equation}
\label{double_statpartial_2}
      \Gamma_1^{\mathrm{fit}} =0.346348\pm0.343585\ ,\qquad
      \Gamma_2^{\mathrm{fit}} =0.739220\pm0.516272\ .
\end{equation}
Although the third resonance is too wide and far away from the point $E_0$, we
made an attempt to locate it using the same optimal expansion parameters. What
we found,
$$
   \mathcal{E}_3^{\mathrm{fit}} = 8.857247-\frac{i}{2}2.847466\ ,
$$
significantly underestimates the width. This is not surprizing, of course. The
truncated series (\ref{approxAB}) can only be accurate within certain circle
around $E_0$. In order to recover the third resonance one has to either take
more terms in the series (\ref{approxAB}) or shift the point $E_0$ (the center
of expansion) down in the complex plane. In both cases the number of fitting
parameters would increase.

\section{Conclusion}
\label{sec.conclusion}
The proposed method is based on proper analytic structure of the parametrized
$S$-matrix that is used to fit experimental data.
 This mathematical correctness
guarantees that after fitting the data, we obtain the $S$-matrix
that is valid in all the channels, even in those where no data are available.
This means that we can obtain the cross section for the channels, which are
experimentally inaccessible, using the data in the other channels.\\

The $S$-matrix properly fitted to the data for real
energies is also valid at
the nearby complex energies. This enables us to extract the resonance
parameters as the real and imaginary parts of the zeros of the
Jost matrix
determinant that coincide with the $S$-matrix poles. In addition to the total $\Gamma$, we are able to rather accurately obtain the channel partial widths.\\

The most important limitation of the method described in this paper, is
the fact that in its present form the method is only applicable to the systems
with short-range interaction forces. A rigorous extension of the method that would include the Coulomb forces, could be done in a way similar to the one described in Ref~\cite{my2011}. This however would
require a modified, much more complicated expression for the
Jost matrix where
all the non-analytic factors (square-root and logarithmic branching points etc.)
are explicitly factorized.\\

The other limitation is the non-relativistic character of the theory used to construct the $S$-matrix. Although the method can still be used in a very wide range of problems dealing with low-energy atomic and molecular collisions, its possible applications in the intermediate- and high-energy particle physics would require relativistic corrections. This could be done in a way that is customary for those who work with mesons, namely, via using relativistic kinematics. For example, the non-relativistic channel momenta (\ref{chmom}) in the Jost matrices (\ref{fin_semi}, \ref{fout_semi}) can be replaced with the corresponding relativistic ones,
$$
   k_n=\frac{1}{\hbar}\sqrt{2\mu_n(E_{\mathrm{kin}}-E_n)+
   \left(\frac{E_{\mathrm{kin}}-E_n}{c}\right)^2}\ ,
$$
where $E_{\mathrm{kin}}=c\sqrt{p^2+\mu^2c^2}-\mu c^2$ is the kinetic (collision) energy.\\

If one accepts the approximate approaches that are traditional for meson-nuclear physics, then using such relativistic momenta in our $S$-matrix that has the correct analytic structure, would be appropriate. Then in a similar simplified fashion the Coulomb corrections could also be introduced as it is done, for example, in Ref.\cite{Arndt}, where the $K$-matrix for a short-range interaction is simply multiplied by a Coulomb-barrier factor $C_0^2(\eta)=2\pi\eta/(e^{2\pi\eta}-1)$. In our case, this trick is equivalent to multiplication of the matrices $A$ and $B$ in Eqs. (\ref{fin_semi},\ref{fout_semi}) by $C_0(\eta)$ and $1/C_0(\eta)$, respectively. The logic behind this is that for charged particles the Riccati-Bessel and Riccati-Neumann functions $j_\ell(kr)$ and $y_\ell(kr)$ are replaced with the corresponding Coulomb functions $F_\ell(\eta,kr)$ and $G_\ell(\eta,kr)$, which at short distances differ from $j_\ell$ and $y_\ell$ by the factors $C_\ell$ and $C_\ell^{-1}$, and the matrices $A$ and $B$ are the factors combining $j_\ell$ and $y_\ell$ into the regular solution (see Eq. (41) of Ref.~\cite{my2011}).
Of course this kind of relativistic and Coulomb corrections are not rigorously justified. However, their usage in meson-nuclear physics is based on some reasonable intuitive argumentation and proved to be working in practical applications.


%
%
\begin{table}
\begin{center}
\begin{tabular}{|c|c|c|c|c|}
\hline
 & $E_r$ & $\Gamma$ & $\Gamma_1$ & $\Gamma_2$\\
\hline
1 & 4.768197 & 0.001420 & 0.000051 & 0.001369\\
\hline
2 & 7.241200 & 1.511912 & 0.363508 & 1.148404\\
\hline
3 & 8.171217 & 6.508332 & 1.596520 & 4.911812\\
\hline
\end{tabular}
\end{center}
\caption{
The exact resonance energies and widths of the
first three resonances generated by the
potential~\protect(\ref{2channelV}). $\Gamma_1$ and $\Gamma_2$ are the partial
widths for the decays into the first and the second channel, respectively.
}
\label{table.2spectrum}
\end{table}

%
%


\begin{thebibliography}{99}
\bibitem{Kukulin}
V. I. Kukulin, V. M. Krasnopolsky, and J. Hor$\acute{\rm a}\check{\rm c}$ek,
{\it ''Theory of Resonances''}, Kluwer Academic Publishers, Dordrecht/Boston/London (1989).

\bibitem{r:ElRakit04}
S.A. Rakityansky, N. Elander,
{\it ''Analyzing the contribution of individual resonance
poles of the $S$-matrix to the two-channel scattering''.}
Int. J. Quantum Chem.  {\bf 109}, 1105 (2006).

\bibitem{r:KsenIJQC09}
K. Shilyaeva, N. Elander, E. Yarevsky,
{\it '' The role of resonances in building cross sections: The Mittag-Leffler expansion in a two-channel scattering.''}
Int. J. Quantum Chem. {\bf 109}, 414 (2009).

\bibitem{r:jpb09}
K. Shilyaeva, N. Elander, E. Yarevsky,
{\it ''Identifying resonance structures in a
 scattering cross section
using the $N^{3+} + H \to NH^{3+} \to N^{2+} + H^+ $ reaction as an example.''}
 J. Phys. B. {\bf 42} ,044011 (2009).

\bibitem{ElandRakit11}
N. Elander and S.A. Rakitiansky,
{\it ''Resonances and their relations to Spectral Densities and Scattering Cross
Sections in the Schr{\"o}dinger formulation''}, Few-Body Systems, On line April27, (2012).

\bibitem{Nichitiu1}
F.Nichitiu,
{\it ''Methods for determining resonances in phase-shift analysis''},
Sov. J. Part. Nucl. {\bf 12} (4), 321 (1981).

\bibitem{Nichitiu2}
F.Nichitiu,
{\it'' Analiza de faza in fizica interactiolor nucleare''}, (in Romanian),
Editura Academiei RSR, Bucuresti (1980).

\bibitem{r:BW36}
G.Breit an E. Wigner,
{\it ''Capture of Slow Neutrons''}, Phys. Rev. {\bf 49}, 519 ( 1936).

\bibitem{r:Fanoprof}
U. Fano,
{\it ''Effects of Configuration Interaction on Intensities and Phase Shifts''},
Phys. Rev. {\bf 124}, 1866 (1961).

\bibitem{Lee}
T.-S. H. Lee,
{\it ''Models for extracting $N^*$ parameters from meson-baryon reactions''},
in: Nstar 2005: Proceedings of the Workshop on the Physics of Excited Nucleons,
World Scientific Pub Co Inc, p.1 (2006).

\bibitem{Tiator}
L. Tiator, S. Kamalov,
{\it ''MAID analysis techniques''},
in: Nstar 2005: Proceedings of the Workshop on the Physics of Excited Nucleons,
World Scientific Pub Co Inc, p.16 (2006).

\bibitem{Svarc}
M. Had$\check{z}$imehmedovi\'c, S. Ceci, A. $\check{S}$varc, H. Osmanovi\'c, and J. Stahov,
{\it ''Poles, the only true resonant-state signals, are extracted from a worldwide collection of partial wave amplitudes using only one, well controlled
pole-extraction method''},
arXiv:1103.2653v1 [hep-ph] (2011).

\bibitem{Badalyan}
A. M. Badalyan, L. P. Kok, M. I. Polikarpov, Yu. A. Simonov,
{\it '' Resonances in coupled channels in nuclear and particle physics''},
Phys.Rep. {\bf 82(2)}, 31 (1982).

\bibitem{Surovtsev}
Yu. S. Surovtsev et al.,
{\it ''Parameters of scalar resonances from the combined analysis of data on processes $\pi\pi\to\pi\pi,\,K\bar{K},\,\eta\eta$ and $J/\psi$ decays''},
arXiv:1207.6937v1 [hep-ph] (2012).

\bibitem{my2009}
           S. A. Rakityansky, N. Elander,
           "{\it Generalized effective-range expansion}",
           J. Phys. A: Math. Theor. {\bf 42},  225302 (2009)

\bibitem{my2011}
           S. A. Rakityansky, N. Elander,
           "{\it Multi-channel analog of the effective-range expansion}",
           J. Phys. A: Math. Theor. {\bf 44},  115303 (2011)

\bibitem{brand} L. Brand,
                "{\it Differential and Difference Equations}",
                 John Wiley \& Sons, Inc., New York, 1966.

\bibitem{Taylorbook}
           J. R. Taylor, ``{\it Scattering Theory}'', John Wiley \& Sons,
           New York, 1972.

\bibitem{my04}
        S. A. Rakityansky, S. A. Sofianos,
        "{\it Jost function for coupled partial waves}",
        Journal of Physics, {\bf A31}, pp. 5149-5175 (1998).

\bibitem{my06}
        S. A. Rakityansky, S. A. Sofianos,
        "{\it Jost function for coupled channels}",
        Few-Body Systems Suppl., {\bf 10}, pp. 93-96 (1999).

\bibitem{bainpot}
        R.A. Bain, J.N. Bardsley, P.R. Junker, C.V.J. Sukumar,
        "{\it Complex coordinate studies of resonant electron-atom
        scattering}",
        J. Phys. B: Atom. Mol. Phys., {\bf 7}, 2189 (1974).

\bibitem{my02}
        S. A. Sofianos and S. A. Rakityansky,
        "{\it Exact method for locating potential resonances and Regge
        trajectories}",
        J. Phys. A: Math. Gen. {\bf 30} 3725 (1997).

\bibitem{minuit1}
           F. James and M. Roos,
           "{\it MINUIT - a system for function minimization and analysis of
           the parameter errors and correlations}",
            Comp.Phys.Comm., {\bf 10}, 343 (1975).

\bibitem{minuit2}
           http://hep.fi.infn.it/minuit.pdf

\bibitem{norotaylor}
        T. Noro, H. S. Taylor,
        "{\it Resonance partial widths and partial photodetachment rate using
        the rotated-coordinate method}",
        J. Phys. B: Atom. Mol. Phys., {\bf 13}, L377 (1980).

\bibitem{Arndt}
        R. A. Arndt, J. M. Ford, L. D. Roper,
        "{\it Pion-nucleon partial-wave analysis to 1100 MeV}",
        Phys.Rev., {\bf D32}, 1085 (1985).
\end{thebibliography}
\end{document}